# Tailoring the optoelectronic, photo-catalytic, thermoelectric and thermodynamic properties of halides Li$_2$InBiX$_6$ (X=Cl, Br, I) for energy conversion - DFT study

Huda A. Alburaih[a], Sikander Azam[b*], N.A. Noor[c*], A. Laref[d], Sohail Mumtaz[e]

[a] Department of Physics, College of Science, Princess Nourah Bint Abdulrahman University, P.O. Box 84428, Riyadh 11671, Saudi Arabia.

[b] University of West Bohemia, New Technologies – Research Centre, 8 Univerzitní, 306 14, Plzeň, Czech Republic

[c] Department of Physics, University of Sargodha, 40100, Sargodha, Pakistan

[d] Department of Physics and Astronomy, College of Science, King Saud University, Riyadh, 11451, King Saudi Arabia

[e] Department of Chemical and Biological Engineering, Gachon University, 1342 Seongnamdaero, Sujeong-gu, Seongnam-si 13120, Republic of Korea

*Corresponding authors: Sikander Azam (Email: azam@ntc.zcu.cz), N.A. Noor (Email: naveedcssp@gmail.com)

**Abstract**

Double perovskite halides are identified as promising materials for diverse applications specifically as renewable energy sources (solar cell devices) and thereby addressing the global energy shortage. This study aims to assess the physical (electronic, optical, dielectric, and thermoelectric) properties of Li$_2$InBiX$_6$ (X = Cl, Br, I) halides by means of density functional approach. Owing to the materials' minimal formation energy, the structures demonstrate stability in the cubic phase. The analyzed double perovskite halides display a semiconducting nature, characterized by a direct bandgap, i.e. Li$_2$InBiCl$_6$ (Eg = 1.7 eV), Li$_2$InBiBr$_6$ (Eg = 1.3 eV), and Li$_2$InAuI$_6$ (Eg = 1.1 eV). Moreover, complex dielectric functions are examined to better understand the optical characteristics of the analyzed halides. These halides are well-suited for optoelectronic applications, as evidenced by the estimated optical parameters that indicate maximum light absorption in IR and visible spectrums. The study extends for analyzing ZT value, Seebeck factor, and electronic conductance within the temperature range 30 – 800 K. The relatively small band gap suggests enhanced suitability for thermoelectric applications, as evidenced by a higher power factor. The photo-catalytic study revealed that Li$_2$InBiX$_6$ are good candidates for the oxidization of H$_2$O at pH values 0-7. Based on the anticipated thermoelectric and optical characteristics, the studied double perovskite halides emerge as potential candidates for energy conversion systems.

## 1. Introduction

Halogen based solar cells (PSCs) have garnered substantial consideration during the recent years due to their remarkable increase in power conversion efficacy (PCE) [1-3]. Nevertheless, due to lacking of long term viability and nontoxic trait, Pb-based DPs continue to impede their widespread commercial use. Recent advancements in enhancing perovskite stability involve the partial or complete substitution of Cs at A-lattice points, contributing to progress by inhibiting the highly volatile constituents [4-6]. Still, the persistent thought regarding the poisonous nature of Pb is still a major obstacle to the global availability of perovskite photovoltaics. As a result, the investigation of inorganic double perovskites (DPs) has been introduced as potential alternative for developing ecofriendly and stable materials suitable for commercializing perovskite-based photovoltaics.

In the 1970s, the first research on DP halides with the chemical composition $A_2B+B^{3+}X_6$ (where X denotes an anionic halogen anion) was conducted. Still, their potential usage in photovoltaics has recently gained significant scientific interest. The efficiency of these photovoltaic energy sources as well as solar energy cells founded on DP halides, is improved by enforcing two critical requirements. First, a small bandgap within the range of (0.8 – 2.2) eV is needed [7], while better light conversion efficiency requires excellent light absorber having energy band gap in the spectrums of (1.3 – 1.7) eV [8, 9]. Consequently, the decrease in energy band gap caused by an increase in ionic radius halide ions can produce a much better bandgap tuning [10]. In this regard, recent studies on both Bi-based and Bi-free DPs have been conducted, most of which have demonstrated that they contain a broad range and indirect band gap [5, 11].

To tackle the above-mentioned challenges, researchers have endeavored to develop free-Pb perovskite materials. This goal has been accomplished through two primary methods: advancements in manufacturing techniques and the incorporation of mixed cations. The manipulation of optoelectronic and magneto-electronic properties can be achieved by introducing cations and anions within $AMX_3$ perovskite materials along with their derivatives, $A_2MX_6$ and $A_2BMX_6$. Notably, $A_2BMX_6$ double-perovskite compounds have demonstrated both non-toxicity and stability as well as bandgap values surpassing 2 eV [12, 13].

For instance, it has been suggested that $Cs_2InBiCl_6$ is the most suited material for photovoltaic absorbers [14]. Photovoltaic devices can benefit from In-based DP halides. A careful examination of the published literature also indicates a limited comprehensive examination of $Cs_2InBiBr_6$ and $Cs_2InBiI_6$ double perovskites [15]. Lead poisoning is a serious drawback, even though lead-based DPs are considered the best for solar applications [16, 17]. Fatima et al. reported the theoretical investigation of $Cs_2InSbX_6$ (X = Cl, Br, I) in terms of its physical durability [18]. Despite this, it is crucial to fully understand the materials' physicochemical properties to determine their performance when utilized in sustainable energy devices. The employment of mBJ functional on DPs ($Rb_2AgSbCl_6$ and $Rb_2AgSbBr_6$) led to the calculations of their spectral features [19, 20]. Mustafa et al. [21], theoretically explored the crystal structure, mechanical, optoelectronic as well as thermodynamic characteristics of $K_2ScAuX_6$ (X = Cl, Br) lead-free DPs. The theoretical investigation of $AeVH_3$ (Ae = Be, Mg. Ca, Sr) observed that the perovskite hydrides also have remarkable transportation and power applications [22].

This study aims to examine the stability and non-toxic nature of $Li_2InBiX_6$ (X = Cl, Br, I) perovskites. The electronic, optical, dielectric, and thermoelectric properties are comprehensively analyzed by means of DFT. The BoltzTraP code is used to compute the thermoelectric characteristics. Since there is a lack of existing literature on this specific research, this theoretical study aims to offer valuable insights to experimental researchers. The findings can be crucial in advancing and propelling the optoelectronic industry as it contributes significant information without prior related studies.

## 2. Computational detail

The computational study of $Li_2InBiX_6$ (X = Cl, Br, I) has been performed using the code Wien2K, an implementation of the DFT framework and based on the FP-LAPW approach [23, 24]. Initially, the optimization of structural parameters was conducted using the PBEsol-GGA functional to acquire the optimized lattice structures [25]. Recognizing the tendency of GGA to underestimate energy bands, we subsequently employed the mBJ approach to ensure more accurate band gap values [19]. The ultimate electronic structure was predicted in the muffin-tin sphere, characterized by a sphere-shaped periodic framework, while the interstitial area exhibited a plane wave-like formation. For this computation, we established $K_{max} \times R_{MT}=8$, $G_{max}=8$, and assigned an $\ell = 10$ in the reciprocal lattice to refine the initial geometry. To ensure accuracy, a k-mesh with 2000 k-points, corresponding to a 12×12×12 grid, was employed for integrating the Brillouin zone [26]. The charge convergence was taken as 0.01 mRy. Additionally, the transport characteristics were calculated using the BoltzTraP code [27], which is established by the traditional Boltzmann transport theory [28], while accounting for the improved electrical structures via mBJ and the convergent energy.

## 3. Results and Discussion

### 3.1 Structural stability

The stability of structural configurations has been examined using cubic structures with the space group (225) Fm-3m. Figure 1(a, b) illustrates the atomic positions of $Li_2InBiX_6$ (X = Cl, Br, I). The octahedra $[InX_6]$ and $[BiX_6]$, isolated by 12-fold coordination, exhibit a decrease in the lengths of In-X and Bi-X bonds within $Li_2InBiX_6$ [29]. Both In and Bi atoms reside at the center of the octahedrons as shown in Figure 1(b). The Wyckoff positions of In / Bi (4a), Li (8c), and X (24e) are specified with the respective coordinates (0, 0, 0), (1/4, 1/4, 1/4), and (x, 0, 0).

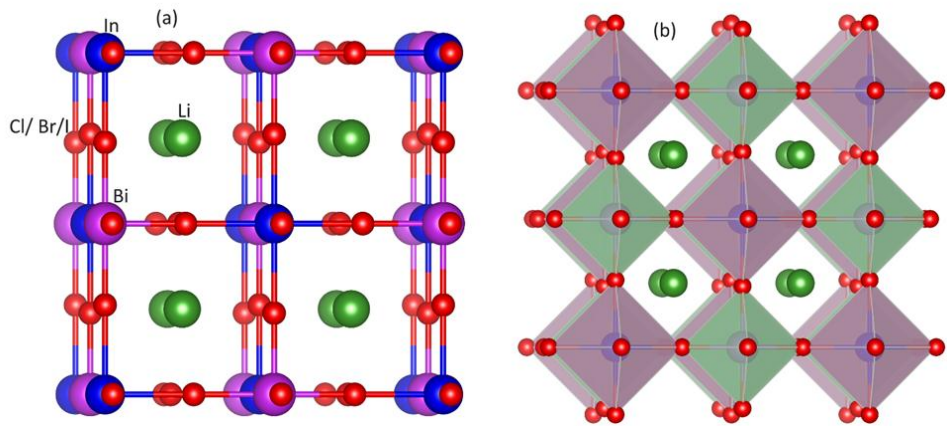

**Figure 1.** Crystal structure plot of double perovskite (a) ball and (b) polyhedral representations for $Li_2InBiX_6$ (X = Cl, Br, I) halides.

PBEsol-GGA functional has been utilized for the optimal structural calculations of $Li_2InBiX_6$ (X = Cl, Br, I), via Murnaghan formula, as shown in Figure 2 [30]. The stable configuration is achieved by the optimization of lattice parameters, bulk modulus and released energy. A decrease in the lattice parameter is observed from Cl (11.19 Å) to Br (11.72 Å) and I (12.52 Å), as referred to in Table 1. An analogous trend was also reported in $Rb_2AgSbCl_6$ (10.76 Å) and $Rb_2AgSbBr_6$ (11.26 Å) [18]. Similarly, the $Cs_2InBiX_6$ (X= Cl, Br, I) exhibited the same increasing trend in $a_0$ values with the rise in atomic radii. The Bulk modulus values fall as the unit cell volume rises (similar to $a_0$), indicating that the $Li_2InBiX_6$ (X = Cl, Br, I) system is compressed appreciably. Consequently, $a_0$ manifests a converse trend to that of B, $B_0 \propto \frac{1}{V_0}$, where Vo indicates the crystals' volume. Because of its favorable ion transport environment, $Li_2InBiX_6$ (X = Cl, Br, I) is a desirable material for energy conversion applications.

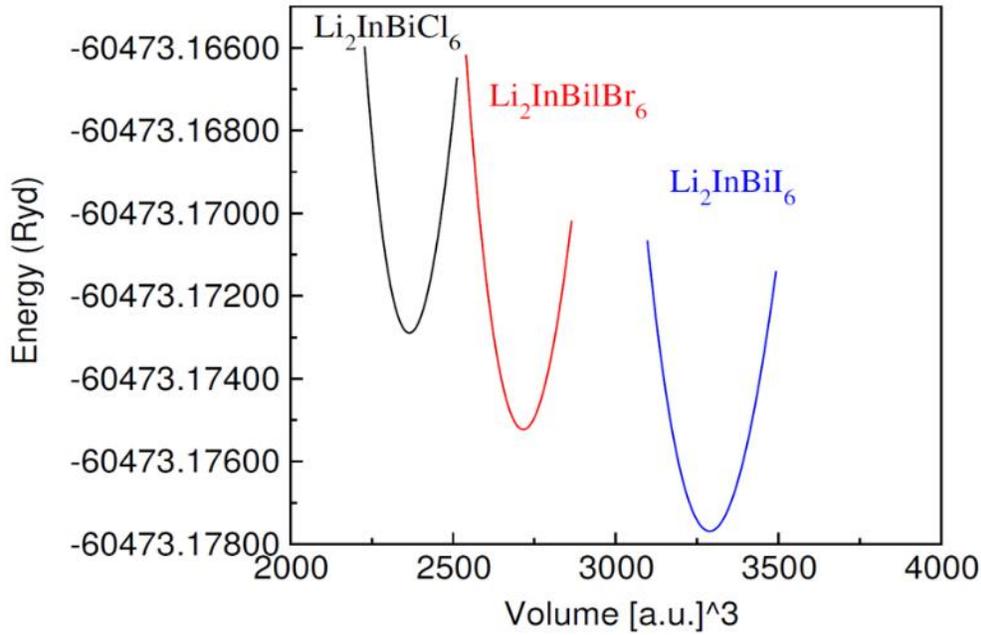

**Figure 2.** Energy-volume optimization plot of $Li_2InBiX_6$ (X = Cl, Br, I) DP halides.

Different physical characteristics displayed by crystal structures can be modified by varying the lattice constant. Such fluctuations may result from altering cations or anions [31]. Since valence electrons are loosely bonded and require small energy to travel towards conduction bands, the lattice constant increases as the ionic radius rises while the band gap reduces by changing the anion from Cl to I. Additionally, by calculating formation energy, the chemical equation below reveals the stability of the studied quaternary DPs related compounds:

$$\Delta E_f = E_{(Li_2InBiX_6)} - [2E_{Li} + E_{In} + E_{Bi} + 6E_X] \qquad (1)$$

Here, $E_{TOT}$ of the calculated halides is denoted as $E_{(Li_2InBiX_6)}$, whereas the isolated atoms energies of Li, In, Bi and Cl/Br are indicated by $E_{Li}$, $E_{In}$, $E_{Bi}$ and $E_X$, respectively. Due to their low formation energy, these compounds can be synthesized since they are stable energetically. A comprehensive analysis of $REGa_3$ (RE = Sc or Lu) composites reported their thermodynamic characteristics [32]. Having the least formation energy, a cubic phase appears as the most stable, with values ranging from -1.97, -1.68 and -1.23 eV for $Li_2InBiCl_6$, $Li_2InBiBr_6$ and $Li_2InBiI_6$, respectively.

The mechanical stability of the compound can be better understood by calculating the elastic parameters. The mechanical parameters ($C_{11}$, $C_{12}$, and $C_{44}$) are used to analyze various parameters as described in Table 1. A minute strain has been applied to deform the structure and subsequently calculate these constants. Comparing its energy value with the equilibrium condition that satisfied the Born-Huang criteria for stability, $C_{44}>0$, $C_{11}-C_{12}>0$, $C_{11}+2C_{12}>0$, and $C_{12}<B<C_{11}$, where B is the Bulk modulus [33]. The computed elastic constants indicates that each compound $Li_2InBiX_6$ (X = Cl, Br, I) has different mechanical stability. Table 1 reports all computed mechanical parameters. A similar analysis of $RBRh_3$ (R = Sc, Y and La) compounds reports that these materials have relatively high elastic constants as compared to ours. So they are ductile, very stiffer and isotropic materials, whereas, ours are nearly ductile, soft and anisotropic in nature [34].

**Table 1.** Calculated values of lattice constant $a_o$ (Å), bulk modulus $B_o$ (GPa), Enthalpy of formation ($\Delta H_f$) and elastic parameters of $Li_2InBiX_6$ (X = Cl, Br, I) halides.

| Parameters | $Li_2InBiCl_6$ | $Li_2InBiBr_6$ | $Li_2InBiI_6$ |
|---|---|---|---|
| | PBEsol-GGA | PBEsol-GGA | PBEsol-GGA |
| $a_o$ (Å) | 11.19 | 11.72 | 12.52 |
| $B_o$ (GPa) | 22.50 | 19.25 | 16.45 |
| $\Delta H_f$ (eV) | -1.97 | -1.68 | -1.23 |

The equations utilized for the calculation of the elastic parameters given in Figure 3(a-d) are defined as follow:

$$B = (C_{11}+2C_{12})/ 3 \tag{2}$$

$$G = 0.5[0.2(C_{11}-C_{12}+3C_{44}) + 5C_{44} (C_{11}-C_{12})/ (4C_{44}+3C_{11}-3C_{12})] \tag{3}$$

$$Y = 9BG/ (3B-G) \tag{4}$$

$$\upsilon = (3B-Y)/ 6B \tag{5}$$

$$A = 2C_{44}/ (C_{11}-C_{12}) \tag{6}$$

The macroscopic properties have been determined by means of DFT calculations, which help in turns to compute DPs halide's atomic position and mechanical properties. Second-order elastic constants are required to better understand the mechanical characteristics of materials [35].

Furthermore, ELAM has been used for visualizing and plotting the compound [36] bulk modulus B, shear modulus G, Young's modulus E, and Poisson's ratio v obtained by applying the Voigt–Reuss–Hill approximations, and the obtained results are presented in Figure 3b. It is found that the Pugh's value increases as function of X, i.e. 2.53, 2.83, and 2.85 for $Li_2InBiX_6$ (X = Cl, Br, I). Pugh's criteria (B/G) indicates whether the material possesses brittle or ductile character [21]. It is determined from the Pugh's criteria that the aforementioned materials are temperature resilient and ductile in nature. For malleable metal-based compounds, the +ve Cauchy value is an indicator of composites' brittleness whereas the -ve value corresponds to their ductility [37]. According to Figure 3d, the materials exhibit ductile nature with positive Cauchy's value indicating their metallic nature. The standard criterion of Poisson ratio will verify and categorize the material as covalent ($\upsilon <0.1$), ionic ($\upsilon <0.25$), or metallic ($\upsilon = 0.33$). According to Poisson index, since its value 0.325, 0.340 and 0.345 for $Li_2InBiX_6$ (X = Cl, Br, I) respectively, confirms its metallic nature with high resilience to volume change.

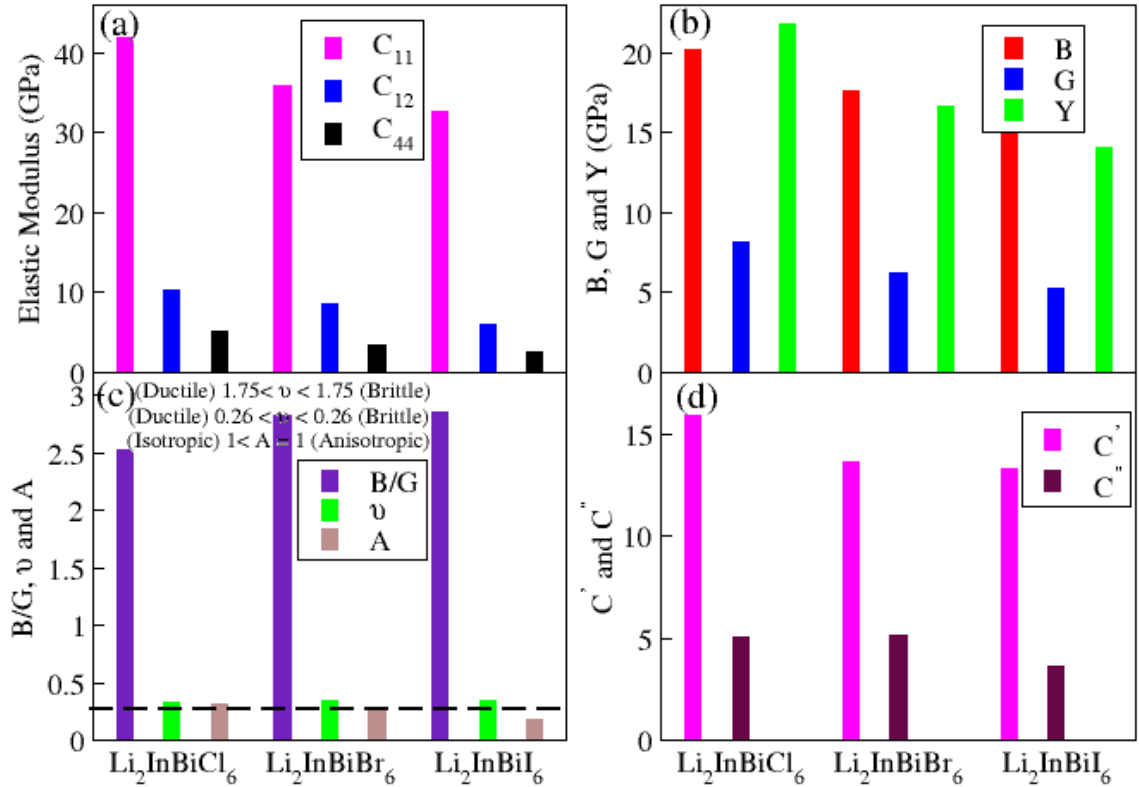

**Figure 3.** Elastic and Mechanical properties of of $Li_2InBiX_6$ (X = Cl, Br, I) DP halides.

For a simple cube lattice, A=1, where A is the Zener Anisotropy ratio, indicates its anisotropic behavior [38], Figure 3c, shows the values of A for DP halides $Li_2InBiX$ (X = Cl, Br, I). Consequently, the calculated values of A (0.32, 0.25 and 0.18) exhibits shear stiffness and because these materials have directional thermal conductivity, so have thermal durability in low temperature environments.

### 3.2 Opto-electronic Properties

The structural information is then acquired by converging the energy utilizing the Trans Blaha-mBJ potential. In general, using the Trans Blaha-mBJ potential yields to a calculated band gap compatible with the experimentally values [39, 40]. Figure 4 illustrated the determined band structure corresponding to $Li_2InBiCl_6$, $Li_2InBiBr_6$, and $Li_2InBiI_6$ DP halides. For all DP halides, the difference of energy among VB maxima and CB minima, called the band gap $E_g$, exhibits a decreasing trend, i.e. 1.7, 1.3, and 1.1 eV for $Li_2InBiCl_6$, $Li_2InBiBr_6$, and $Li_2InBiI_6$, respectively, and each exists at the same Γ-points. This indicates that the DP halides under investigation have a narrow direct $E_g$, an essential characteristic of materials with potential usage in optoelectronic devices. Excluding the lattice thermal contribution and the direct transition of electrons to accessible conduction state(s) from confined valence conditions, the direct band gap nature makes it easier to absorb light.

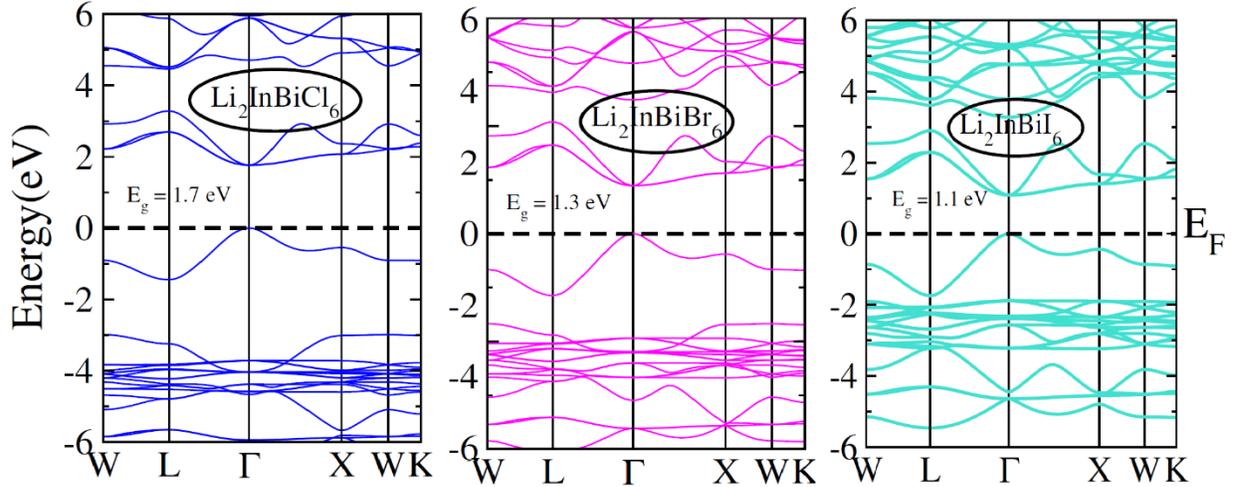

**Figure 4.** Estimated band structure for $Li_2InBiCl_6$, $Li_2InBiBr_6$ and $Li_2InBiI_6$ DP halides.

For such materials having a direct band gap, critical energy values can be accurately calculated to respond to the imaging frequencies. The band gap of $Li_2InBiX_6$ (X = Cl, Br, I) estimated by means of TB-mBJ functional are given in Table 2.

**Table 2.** Computed $E_g$(eV) magnitudes with mBJ functional and optical parameters at 0 energy of $Li_2InBiX_6$ (X = Cl, Br, I).

| Parameters | $Li_2InBiCl_6$ | $Li_2InBiBr_6$ | $Li_2InBiI_6$ |
| --- | --- | --- | --- |
| $E_g$ (eV) | 1.7 | 1.3 | 1.1 |
| $\varepsilon_1$ (0) | 3.2 | 4.1 | 5.2 |
| n (0) | 1.7 | 2.05 | 2.6 |
| R (0) | 0.08 | 0.10 | 0.14 |

The widely employed potential mBJ induces a change in energy states, leading to a calculated band gap that differs from the existing theoretical data. Additionally, it is known that the mBJ potential may compute band gaps that agree with the results of earlier studies. However, no experimental data is available to compare with the investigated DP halides. The experimentally examined $Cs_2InBiX_6$-like DPs (X = Cl, Br, or I) are the only existing data. The $E_g$ values were found to be 2.21, 3.2, and 2.77 eV for $Cs_2AgBiBr_6$ (hydrothermal method), $Cs_2AgBiCl_6$ (solid-state approach) and $Cs_2AgInCl_6$ (sol-gel technique), respectively [41]. An excellent photovoltaic material requires an optimum band gap energy ranging between 0.8 and 2.2 eV [7], which enables it to display maximum energy conversion efficiency. For $Li_2InBiBr_6$, the computed $E_g$ value of 1.3 eV is the best among the other two compounds i.e. $Li_2InBiCl_6$ (1.7 eV) and $Li_2InBiI_6$ (1.1 eV), because it lies in near infrared spectrum (red edge). $E_g$ values of the two compounds also lie in the infrared region, but it is better to tune the band gap effectively by substituting halogen anions (X) requirement. The materials with a small direct band gap under study reflect the characteristics of Pb-based Perovskite's solar devices. The non-toxicity of these materials gives them an edge over Pb-based Perovskites. Hence, the investigated materials are considered more favorable than optimal Pb-based materials [14].

To determine the effect of electrons in multiple orbitals, such as s, p, d, and f, the density of states (DOS) in terms of total density of state (TDOS) and partial density of states (PDOS) have been also computed [42], see Figure 5. The optimum electron involvement for every component has been also examined; Figure 5 illustrates which orbitals participate significantly compared to others: 2s-states of Li, In 5p-states, bismuths' 6p-states, X atoms p-states.

The effectual attraction caused by hybridization shows a low energy shift, resulting in the bottom of conduction band states of In-5p and Bi-6p. Thus, halogen's ionic radius affects the hybridization efficiency of band gap control and offers the potential for designing devices that endure various incident energies. The plot shows that in conduction band main contribution of In-5p, Bi-6p and halogen p-orbital (Cl-3p, Br-4p and I-5p) can be seen, whereas, the band gap size of decreases from largest to smallest for $Li_2InBiCl_6$, $Li_2InBiBr_6$ and $Li_2InBiI_6$ respectively, also all the three materials have a zero DOS region near the $E_f$ confirming their semiconducting nature. The contribution of Li is minimal as seen in the plot, showing that it does not participate in bonding interactions near $E_f$ and it behaves only as charge balancer i.e. to maintain overall charge neutrality.

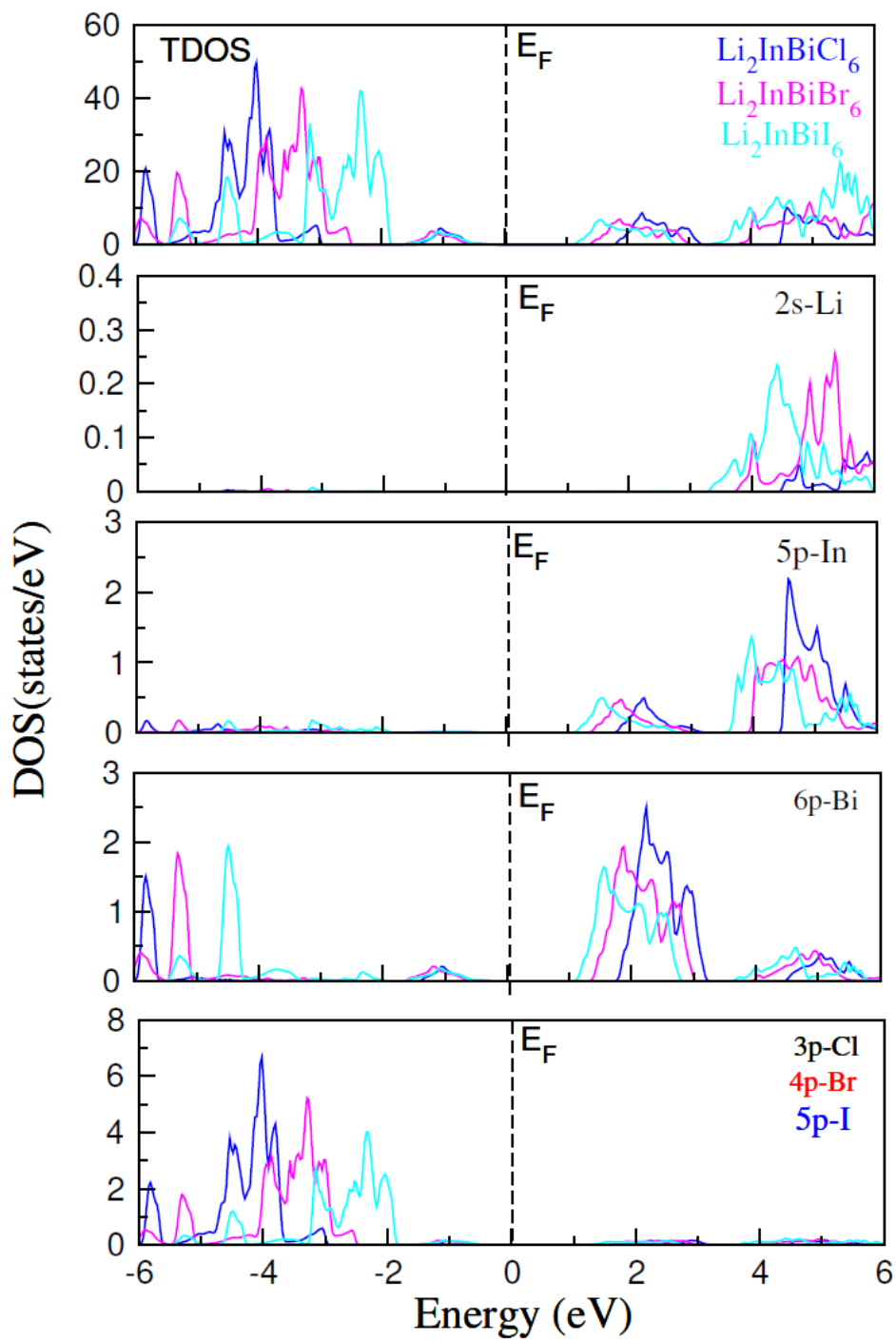

**Figure 5.** Calculated DOS (TDOS and PDOS) of Li$_2$InBiCl$_6$, Li$_2$InBiBr$_6$ and Li$_2$InBiI$_6$ halides.

The dielectric constant $\varepsilon(\omega)$, extinction coefficient $k(\omega)$, refractive index $n(\omega)$, absorption $\alpha(\omega)$ as well as reflectance $R$) are the estimated spectral features of Li$_2$InBiX$_6$ (X=Cl, Br, I), refers to

Figure 6(a-f). The dielectric constant $\varepsilon(\omega)$ is divided into two parts ($\varepsilon_1(\omega)$ and $\varepsilon_2(\omega)$) corresponding to polarization and absorption of material. The mathematical expression of real factor [43] is given as:

$$\varepsilon_1(\omega) = 1 + \frac{2}{\pi} P \int_0^\infty \frac{\omega' \varepsilon_2 \omega'}{\omega'^2 - \omega^2} d\omega' \qquad (7)$$

The converse relation between $\varepsilon(\omega)$ at $\varepsilon_1(0)$ and energy band has been explained by the Penn's Formulation [44]. The greatest resonance rate of energy transmission is maintained at 2.62 eV (Li$_2$InBiCl$_6$), 2.48 eV (Li$_2$InBiBr$_6$), and 0.94 eV (Li$_2$InBiI$_6$). Table 2 reports the constant energy measurements for an actual component of the dielectric equation. It is found that $\varepsilon_1(0)$ of Li$_2$InBiX$_6$ increases (from 3.2 to 5.2) when the energy gap shrinks (from 1.7 to 1.1 eV) due to the transition from Cl to I. The peak swiftly proceeds towards the smallest value upon reaching the resonant point. It is remarkable that the halogen-based DPs under investigation have a cube shaped configuration. The scattering band concept recognizes this framework, and $\varepsilon_2(\omega)$ illustrates the way the bands are precisely corresponding to the imaginary component of the dielectric constant. Moreover, photons with energies under the ensuing energy gaps of Li$_2$InBiCl$_6$ (Eg = 1.7 eV), Li$_2$InBiBr$_6$ (Eg = 1.3 eV), and Li$_2$InBiI$_6$ (Eg = 1.1 eV) have computed coefficient $\varepsilon_2(\omega) = 0$, as shown in Figure 5b. The obtained dielectric coefficients indicate that the halogens are essential elements in each photovoltaic device as they have an optimal attraction for absorbing photons in the visible and infrared portions of the electromagnetic (EM) spectrum. To calculate the phase speed of EM waves at different frequencies through the material, the plot of n($\omega$) is shown in Figure 6c. The K($\omega$) describes the absorption of EM radiations while moving through material.

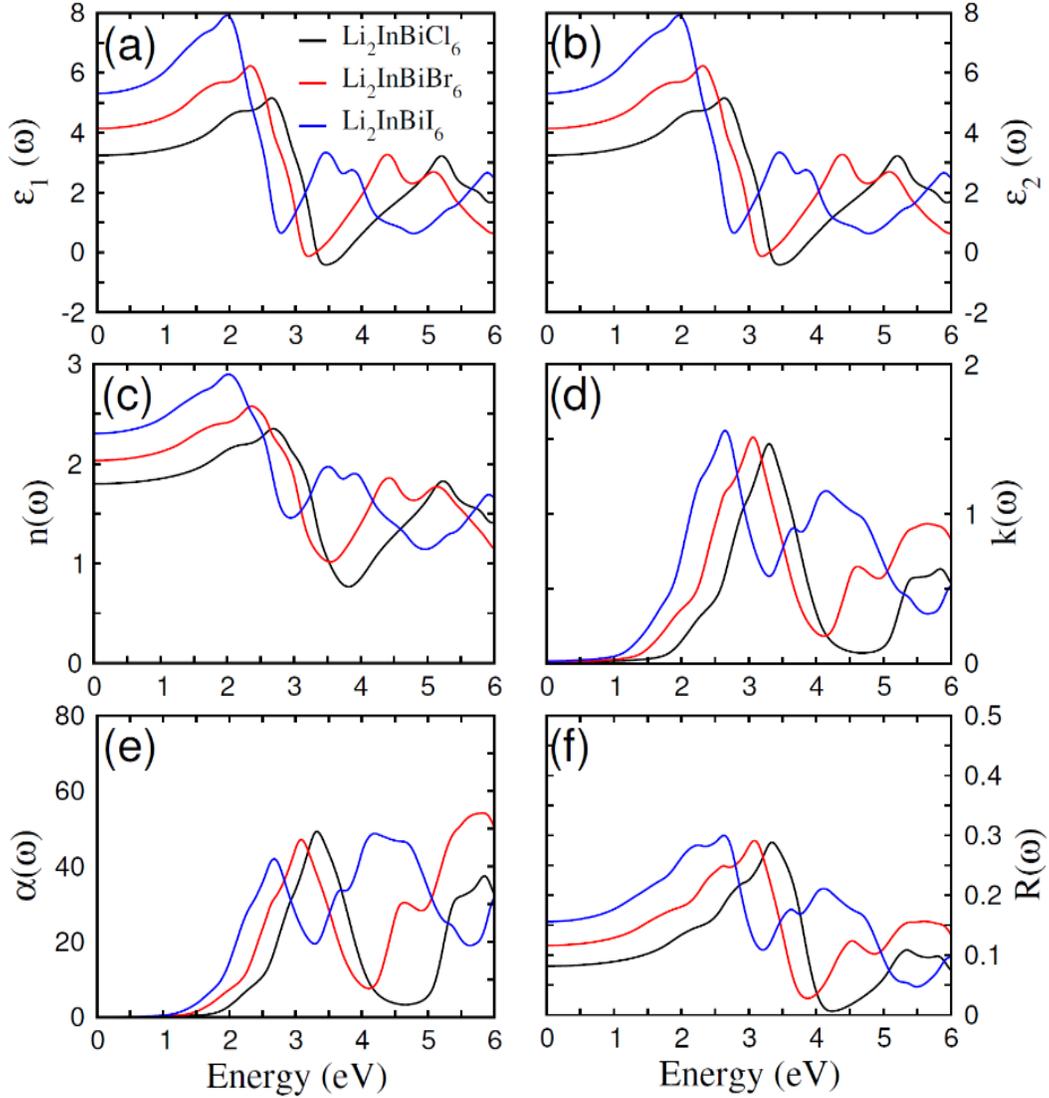

**Figure 6.** The computed (a) real part $\varepsilon_1(\omega)$, (b) imaginary part $\varepsilon_1(\omega)$, (c) refractive index n($\omega$), (d) the extinction coefficient k($\omega$), (e) absorption $\alpha(\omega)$, and (f) reflectivity R($\omega$) of $Li_2InBiCl_6$, $Li_2InBiBr_6$ and $Li_2InBiI_6$ halides.

It is observed that n(0) is 1.7, 2.05 and 2.6 for $Li_2InBiCl_6$ $Li_2InBiBr_6$ and $Li_2InBiI_6$, respectively. This is due to the reason that as the halide atom gets heavier the value of n(0) increases and it is consistent with the decreasing trend in bandgap. The value of $n(\omega)$ at wavelength of 400 nm (i.e. blue region of EM) lies in the range of 1.7 – 4.5, which is a characteristic range of materials suitable for photovoltaic applications. The highest $K(\omega)$ values observed for the composite materials are in the range of 2.5 – 3.5 eV, which corresponds to blue and near ultraviolet region of EM spectrum, indicating strong absorption in this optical range. It is evident that the absorption spectrum $\alpha(\omega)$ begins for Li$_2$InBiCl$_6$, Li$_2$InBiBr$_6$ and Li$_2$InBiI$_6$ at 1.7, 1.3, and 1.1 eV, respectively, and reaches its ultimate values at 3.3, 3.0 and 2.5 eV, as shown in Figure 6e. A rapid rise in the absorbance rate results from the expansion of absorption boundaries due to the incidence of light rays. The findings also reflect an excellent similarity in the high-frequency region, when semiconducting materials' absorbance factor often begins to fall and subsequently rise. In comparison with a similar system Cs$_2$InBiX$_6$ (X = Cl, Br, I), the researched IR absorption of halogen-based DPs has been less explored [46]. In the energy spectrum range 2.5 – 3.5 eV, Li$_2$InBiX$_6$ exhibits the highest value of reflectivity R($\omega$), as seen in Figure 6f. It can be concluded that Li$_2$InBiCl$_6$ demonstrates the optimum spectral parameters, and thereby presents potential usage in optical devices.

### 3.3 Thermoelectric properties

Another efficient and economical approach consists in producing electrical energy through heat conversion to meet the world's energy demands. Therefore, the calculations of thermoelectric features are critical to assess materials' properties and performance as potential energy carrier. In recent years, DPs have often been employed in thermoelectric appliances as well as halogen oriented photovoltaic devices [47]. By conducting thermal studies with respect to temperature, Figure 6 indicates that only Li$_2$InBiCl$_6$ displays an appropriate energy gap. The mBJ approach has been employed to compute electrical conductivity ($\sigma/\tau$), Seebeck coefficient (S), thermal conductivity (k$_e$/$\tau$), Specific heat capacity (C$_v$), power factor ($\sigma$S$^2$) and Figure of merit (ZT) of Li$_2$InBiCl$_6$, Li$_2$InBiBr$_6$ and Li$_2$InBiI$_6$ halides.

The most significant component in the predicted thermal characteristics is the electronic conductivity [48, 49]:

$$\sigma_{\alpha\beta}(\varepsilon) = \frac{1}{N}\sum_{i,k} \sigma_{\alpha\beta}(i,k) \frac{\delta(\varepsilon-\varepsilon_{i,k})}{\delta(\varepsilon)} \quad (8)$$

$$\sigma_{\alpha\beta}(i,k) = e^2 \tau_{i,k} v_\alpha(i,\vec{k}) v_\beta(i,\vec{k}) \quad (9)$$

where $N$ = total count of $k$-points, $\tau$ = relaxation time, (i, k) for the group velocity, and $k$ = wave vector. In order to analyze the thermal response, the following additional conditions must be met:

$$\sigma_{\alpha\beta}(\alpha,\mu) = \frac{1}{\Omega} \int \sigma_{\alpha\beta}(\varepsilon) \left[-\frac{\partial f_0(T,\varepsilon,\mu)}{\partial \varepsilon}\right] d\varepsilon \quad (10)$$

$$S_{\alpha\beta}(T,\mu) = \frac{1}{eT\Omega\sigma_{\alpha\beta}(T,\mu)} \int \sigma_{\alpha\beta}(\varepsilon)(\varepsilon-\mu) \left[-\frac{\partial f_0(T,\varepsilon,\mu)}{\partial \varepsilon}\right] d\varepsilon \quad (11)$$

where $\Omega$ is the unit cell volume, $\beta$ the electron charge, $\mu$ the carrier concentration, $\alpha$ the tensor vector, and $f_0$ the Fermi Dirac distribution. The thermoelectric efficiency and output coefficient may be calculated using the below formula:

$$ZT = (\sigma S^2/\kappa)\, T \qquad (12)$$

where S is the Seebeck factor, $\sigma$ the electronic conductivity, $\kappa$ the thermal conductivity, and $\tau$ the relaxation time [50]. A reduction in $\sigma$ and S magnitudes induces a reduction in $\kappa$ value, which corresponds to their converse relation. Besides, $k_e$ and $\sigma$ can be enhanced by $\tau$, as demonstrated by following the traditional Boltzmann thermodynamic equation. Besides, it is found that the constant $\tau$ remains unchanged in the aforementioned algorithm. The calculations of materials' transport features are temperature dependent. All thermoelectric characteristics have been assessed over multiple T spectrums ranging from (300 – 800) K and the results near ambient temperature are noteworthy. Figure 7 (a) shows the association between $\sigma$ and T, for $Li_2InBiCl_6$, an enhancement in $\sigma/\tau$ is computed with the respective findings [$0.026 \times 10^{19}$/$\Omega$ms, $0.024 \times 10^{19}$/$\Omega$ms, and $0.019 \times 10^{19}$/$\Omega$ms] at 200 – 800 K. Consequently, it can be said that owing to its suitable energy gap, $Li_2InBiCl_6$ exhibits maximal conductance throughout elevated temperatures.

One of the important thermal features of compounds is the Seebeck factor [51], described as:

$$S = \Delta V / \Delta T \qquad (13)$$

Figure 7 (b) illustrates the calculated values of S factor in the temperature range 300 – 800 K. The computed values of the Seebeck coefficient at 300K are 220, 230, and 250 µV/K for $Li_2InBiCl_6$, $Li_2InBiCl_6$, and $Li_2InBiCl_6$ respectively. Besides, it can be observed that's its value declines slowly around ambient temperature then remains constant till 800 K. The S factor may be accounted for the estimation of thermal efficiency of heat generators. When it pertains to determine the thermoelectric performance of materials using $k_e$ and $k_l$, thermal conductivity plays a major role. Since the computed electrical properties are more for suitable for direct bandgap materials, it is entirely due to the limits of BoltzTraP algorithm which is not accurate for indirect bandgap materials. Figure 7(c) reflects that the thermal conductivity is also addressed as a critical efficiency measure of material related to temperature [52]. At room temperature, the values of (ke/$\tau$) are computed for $Li_2InBiCl_6$ ($0.12 \times 10^{14}$ W/mK), $Li_2InBiBr_6$ ($0.11 \times 10^{14}$ W/mK), and $Li_2InBiI_6$ ($0.11 \times 10^{14}$ W/mK).

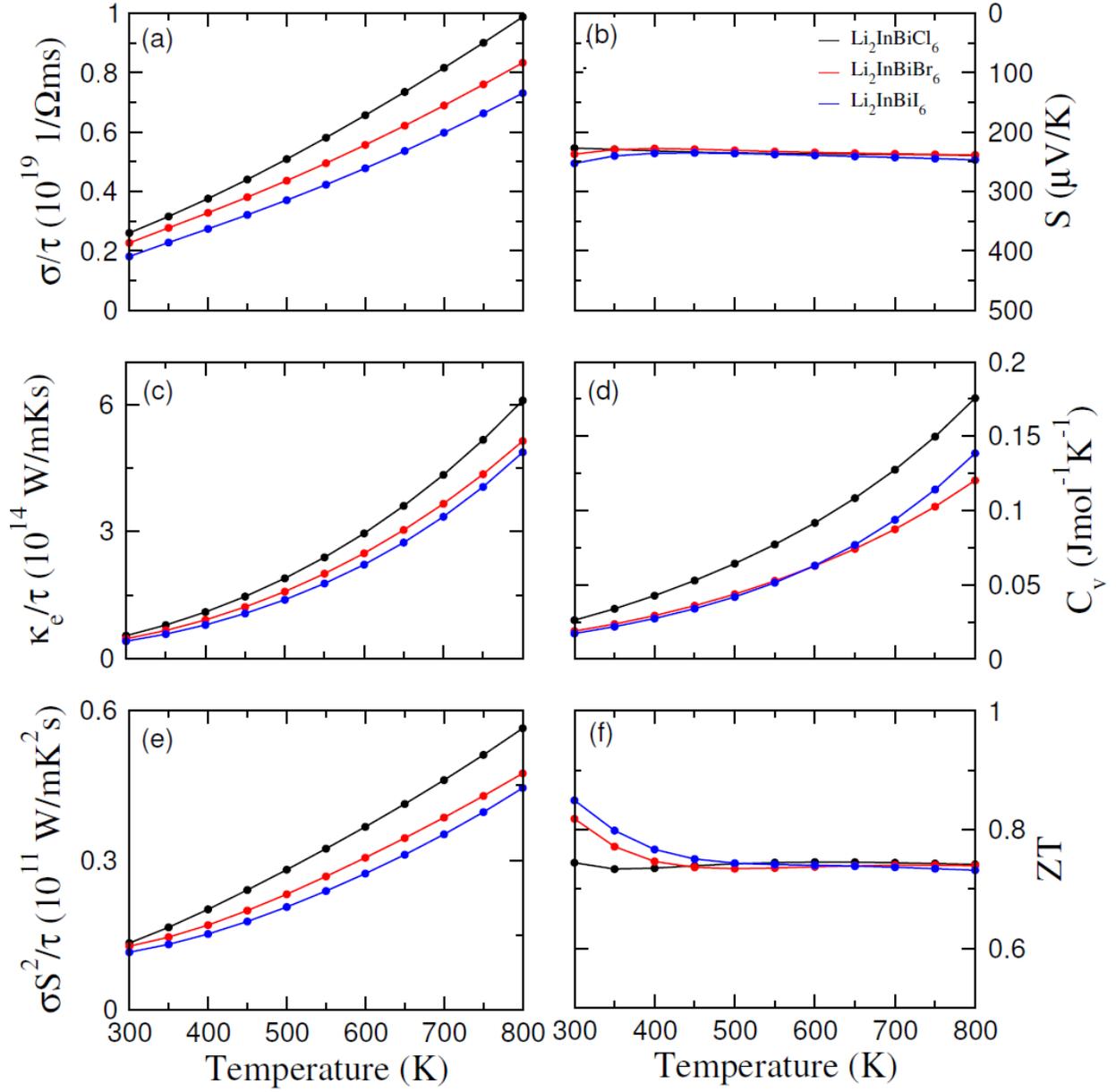

**Figure 7.** The computed (a) electrical conductivity ($\sigma/\tau$), (b) Seebeck coefficient (S), (c) thermal conductivity ($k_e/\tau$), (d) Specific heat capacity ($C_v$), (e) power factor ($\sigma S^2$) and (f) Figure of merit (ZT) of $Li_2InBiCl_6$, $Li_2InBiBr_6$ and $Li_2InBiI_6$ halides.

The PHONOPY code with the finite displacement method has been employed to acquire IFCs-2 [53]. IFCs-3 is mandatory for the computation of lattice $k_e/\tau$ (anharmonic term). The estimation of 3-phonon dispersion rate involves the solution of linearized BTE with the 3$^{rd}$ order IFCs. These IFCs have been computed using a 2×2×1 supercell, and $E_{TOT}$ are attained with point meshes of

2×2×1. Subsequently, the PHONO3PY code is utilized to address the BTE employing a single mode τ approach and calculate the lattice heat conduction [54]. As illustrated in Figure 8, the analyzed lattice heat conduction at 300 K is found to be 1.6, 1.25, and 2.25 Wm$^{-1}$K$^{-1}$ for Li$_2$InBiX$_6$ (X=Cl, Br, I) halides. The value of lattice thermal conductivity decreases with the rise of temperature and its minimum value is observed as (0.65, 0.50, and 0.80) Wm$^{-1}$K$^{-1}$ for Li$_2$InBiX$_6$ (X=Cl, Br, I) halides (see the Figure 8).

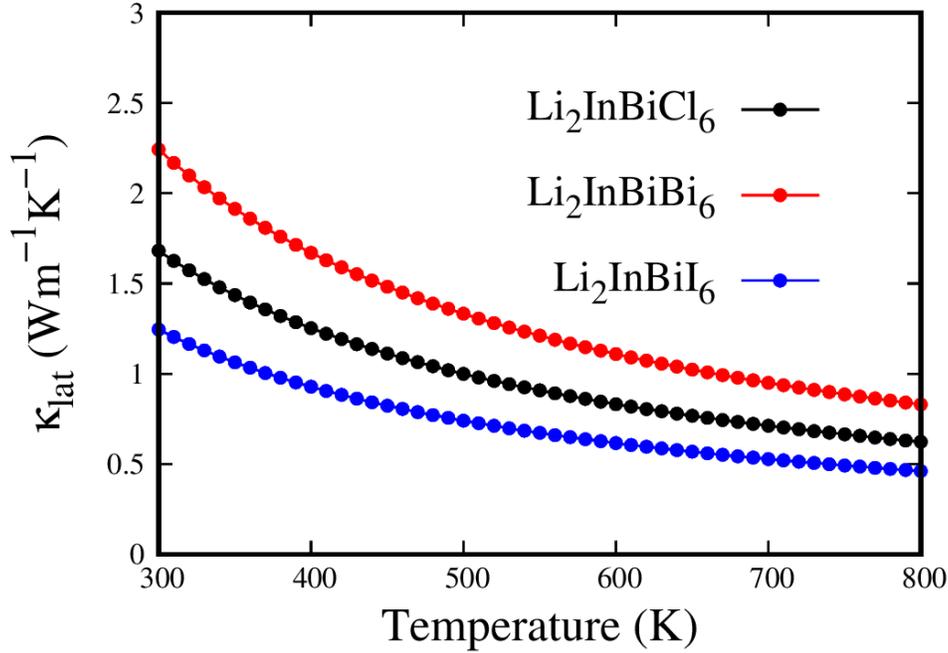

**Figure 8.** The lattice thermal conductivity as a function of temperature of Li$_2$InBiCl$_6$, Li$_2$InBiBr$_6$, and Li$_2$InBiI$_6$ DP halides.

Furthermore, the parameter C$_V$, called the specific heat capacity, represents the ability of a device to conserve heat. Elevated temperatures are typically accompanied by rising heat capacity. The device melts from solid to liquid, then evaporates into vapor or transforms into a gas. Lower temperature range causes a decline in vibrations intensity and resultant harmonic assumptions are typically effective. Each materials' C$_V$ parameter is influenced by temperature. The activation of additional modes of vibrations with increasing temperature (which can be observed in Figure 7d) is the main reason behind higher C$_V$ values for Li$_2$InBiX$_6$ (X= Cl, Br, I).

Based on the above results, Li$_2$InBiX$_6$ can be considered as the best choice for the application with the formula PF= σS^2/τ. The expression σS$^2$ measures the power of a given compound. Figure 6e shows the evolution of PF *vs.* temperature. It is noticed that Li$_2$InBiX$_6$ exhibits larger S magnitudes at ambient temperature, which is consistent with the observed PF values. Moreover, it should be mentioned that an increase in temperature leads to the rise in PF ratio. Therefore, the investigated DP halides have potential utilization because of phenomenal phase stability even under extreme conditions, thanks to the improvement of power factor with the rise of temperature.

Figure 8(f) depicts the computed ZT of Li$_2$InBiX$_6$ (X= Cl, Br, I) at 300K. The association of material's thermal conductivity with the calculated PF and ZT values derived at ambient temperature is found to be 0.74 (Li$_2$InBiCl$_6$), 0.82 (Li$_2$InBiBr$_6$), and 0.85 (Li$_2$InBiI$_6$). The optimal ZT calculations for Li$_2$InBiI$_6$ are also reported in the current study. Figure 6f shows that the ZT values slightly decrease with the temperature increment. Therefore, the potential of implementing DP halides in the applications such as thermoelectric power stations and cooling systems has been improved by the successful conversion of thermal power into electrical power. In fact, heat creation results the energy loss and so significant decrease of thermal conductivity, which further helps to increase the productivity of semiconducting systems by maintaining temperature gradient. [55]. The compounds having the lowest rate of thermal conduction are preferable for photovoltaic devices [56]. Thus, one of the main goals consists in determining the lattice heat transmission accurately.

### 3.4 Photo-catalytic properties

Solar energy can be captured and hydrogen can be produced by dissociating water with the right indirect bandgap semiconductors [57-59]. Thus, clean, sustainable energy can be produced by photocatalytic water splitting [60, 61]. Electrons reduce water during the photocatalytic process, while holes oxidize it [62]. The conduction and valence band photocatalytic water splitting must be greater than the oxidation (reduction) potential of 0 (1.23) eV for this activity for all materials considered [63]. For photocatalytic water splitting, the typical oxidation as well as reduction potentials are -4.44 eV and -5.67 eV, respectively, on the hydrogen scale, as shown in figure 9. The Fermi level is set to (-4.4) eV in order to determine the band edge positions of the VB and CB with respect to standard oxidation [64]. At 0 eV = -4.4 eV and 1.23 eV = 5.67 eV, respectively, the CB and VB are positioned [65]. The figure makes it evident that Li$_2$InBiCl$_6$ showing for both oxidation and reduction of H$_2$O while Li$_2$InBiBr$_6$ and Li$_2$InBiI$_6$ shows good response for both oxidation only. So Li$_2$CuYCl$_6$ is better option for the oxidization and reduction of water.

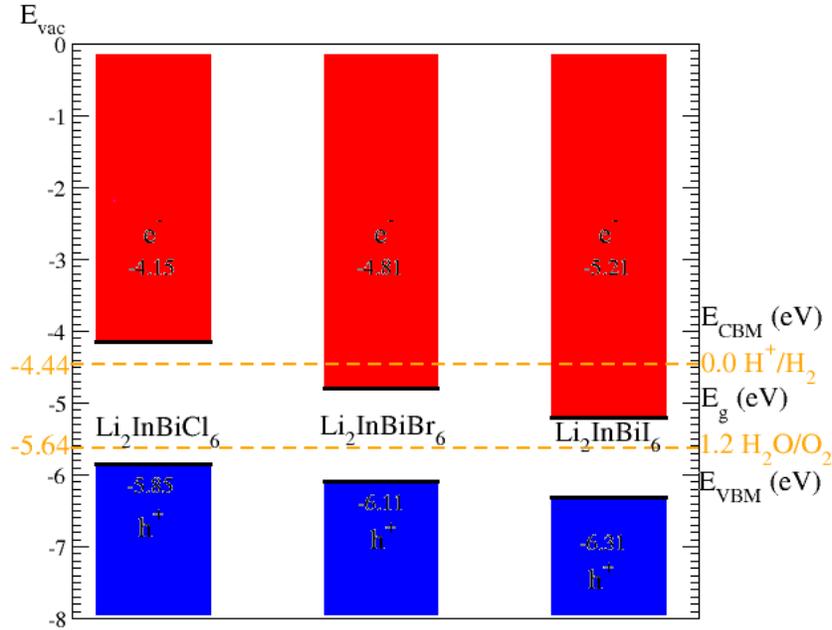

Figure 9: Calculated photocatalytic properties of Li$_2$InBiX$_6$ (X = Cl, Br, I).

### 3.5 Thermodynamic properties:

Figures 10-12 presents the temperature-dependent thermodynamic properties of the compounds Li$_2$InBiCl$_6$ (X=Cl, Br, I) under different hydrostatic pressures (0 GPa, 2 GPa, and 4 GPa). Each subfigure shows how a specific property varies with temperature from 0 to 800 K, and how it is influenced by pressure. In figure 10-12 (a) Entropy S (J/mol·K) vs. Temperature (K) S increases with temperature, as expected. At a given temperature, entropy decreases with increasing pressure (0GPa > 2GPa > 4GPa). Higher pressure reduces lattice vibrations and configurational disorder, leading to lower entropy.

The quasi-harmonic Debye model within the GIBBS2 computational frame is used to explore the thermodynamic behavior of Li$_2$InBiCl$_6$ (X=Cl, Br, I) spinels [66]. Using this approach, we calculated several thermal characteristics such as entropy (S), Debye temperature ($\theta_D$), heat capacity (C$_V$), and thermal expansion coefficient ($\alpha$) have been calculated under the influence of hydrostatic pressures of 0, 2, and 4 GPa and within temperature range of 0–600 K. The mechanical stability, vibrational characteristics and phonon-related behavior of these materials have been revealed through these parameters and that are important for assessing their potential in energy-related applications [67]. The entropy (S) gives valuable insight into the activity of vibrations (phonons) and the level of atomic disorder in the crystal. Figures (10a, 11a & 12a) represent the consistency of third law of thermodynamics as entropy is almost zero at 0 K and increases steadily with temperature [68]. The curves show that entropy rises sharply at lower temperature due to the rapid activation of phonons but at higher temperature this rate slows down as the available vibrational modes become fully occupied. Interestingly, the pressure significantly influences the entropy of these investigated materials at elevated temperatures. The observed entropy is 410 J/mol·K for $Li_2InBiCl_6$, 490 J/mol·K for $Li_2InBiBr_6$ and 520 J/mol·K for $Li_2InBiI_6$ at 300 K and zero pressure (see Fig. 10-12). This increase from Cl to I is due to the greater number of vibrational modes and heavier atomic mass in the I-based composition [69].

The stiffness of the material and how its atoms vibrate reflected through Debye temperature ($\theta_D$) and closely related to the way entropy changes with temperature and pressure. Figures (10-12)b show that $\theta_D$ stays constant at very low temperature but begins to drop steadily after around 200 K due to stronger anharmonic vibrations and phonon softening at higher temperatures [70]. Similar trend of $\theta_D$ has been observed in all compositions by inducing different pressure. The material becomes more ordered and rigid at high pressures due to the observed decrease in entropy and strengthening of phonon vibrations under pressure.

The relationship between Debye temperature ($\theta_D$) and entropy (S) also affects the heat capacity at constant volume (Cp). Figures 10,11 and 12(c) indicate that Cp follows a typical $T^3$ trend at lower temperature and showing agreement with Debye theory. All phonon vibrations become fully active at higher temperature which gradually levels off this Cp and approaches the classical Dulong–Petit limit [71]. The shifting from quantum to classical is more noticeable below 220 K and Cp increases more slowly and eventually stabilizes beyond this point. Pressure slightly reducing Cp by limiting the extent of atomic vibrations under compression and doesn't much effect it [72]. The observed heat capacities (Cp) at zero pressure and 300K are 260 J/mol·K for $Li_2InBiCl_6$ and 265 J/mol·K for $Li_2InBiBr_6$ 270 J/mol·K for $Li_2InBiI_6$. Similar to the pattern seen as that of entropy is reflected through this increase which indicate the influence of atomic mass and crystal structure on it [73]. The thermal expansion coefficient ($\alpha$) explains that how the volume of these materials changes with temperature under different pressures. This $\alpha$ is highly responsive to both temperature and pressure as shown in Figures 10-12(d). Its trend is similar to that of Cp but with an even stronger dependence on pressure. The thermal expansion coefficient ($\alpha$) rises sharply at lower temperature revealing the growing influence of anharmonic (non-linear) atomic vibrations as heat energy builds up. A typical behavior of materials as they approach the classical Dulong–Petit limit is observed with the higher temperature increase and α begins to level off [74]. At 300 K and zero pressure, α is calculated as $10.5 \times 10^{-5}$ $K^{-1}$ for $Li_2InBiCl_6$ and $11.7 \times 10^{-5}$ $K^{-1}$ for $Li_2InBiBr_6$ and $12.5 \times 10^{-5}$ $K^{-1}$ for $Li_2InBiI_6$. The higher value in the iodide composition is due to greater polarizability and larger atomic size of I which make the crystal lattice more responsive to thermal vibrations and less rigid. The thermal expansion coefficient (α) noticeably decreases in the studied materials with the application of hydrostatic pressure and aligns with what is typically seen in condensed materials under the effect of compression [75]. The reduction in α occurs because pressure weakens anharmonic effects, and limits atomic vibrations, making the material less prone to thermal expansion and more rigid.

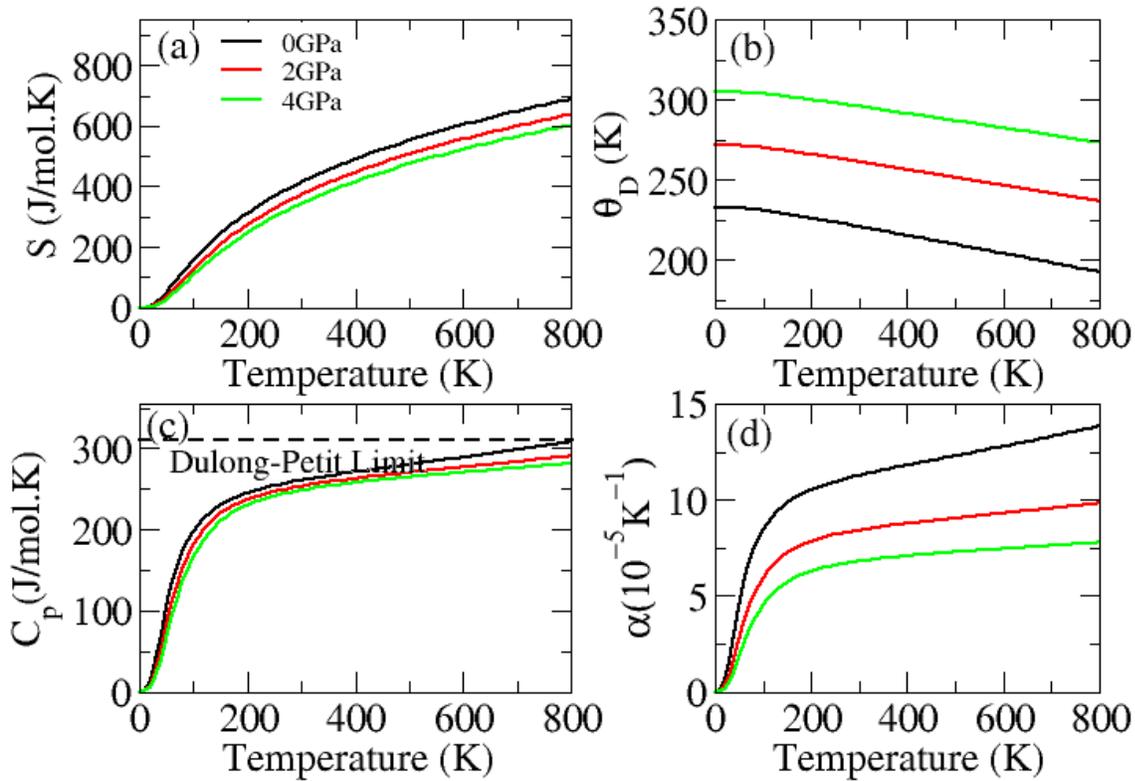

Figure 10: Calculated Thermodynamic properties of $Li_2InBiCl_6$ (a) Entropy vs Temperature (b) Debye Temperature vs Temperature (c) Heat capacity vs Temperature (d) Co-efficient of volume thermal expansion vs Temperature.

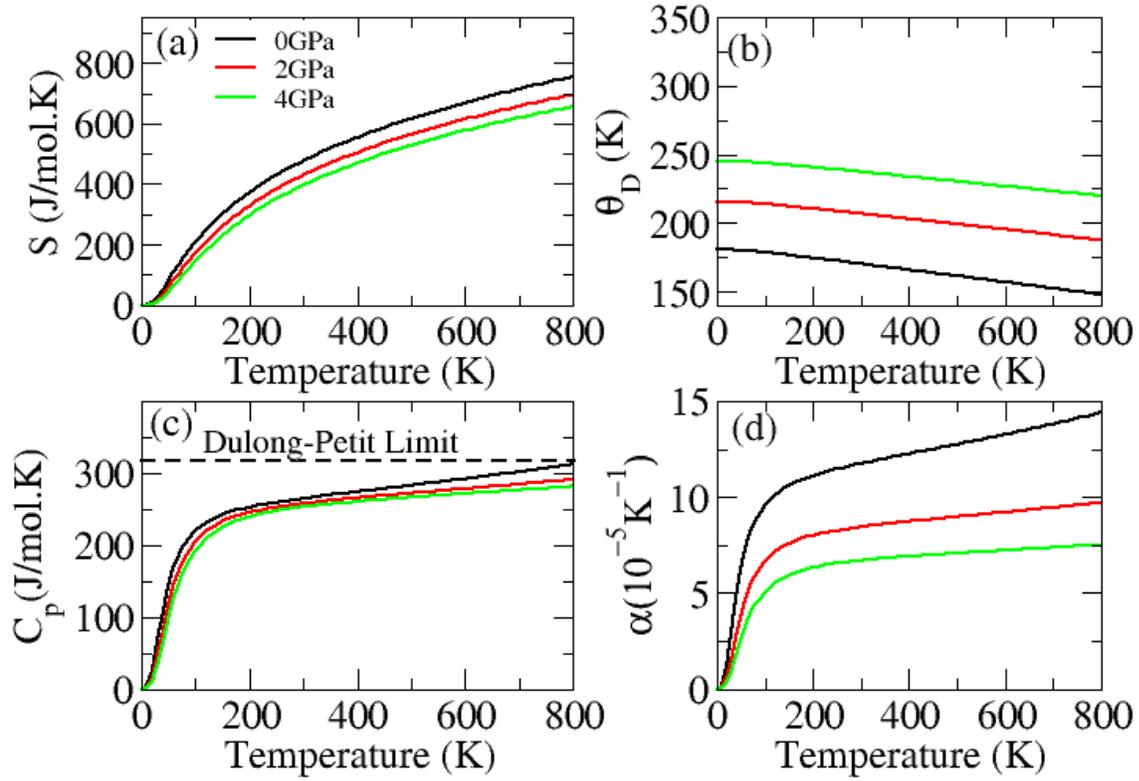

Figure 11: Calculated Thermodynamic properties of $Li_2InBiBr_6$ (a) Entropy vs Temperature (b) Debye Temperature vs Temperature (c) Heat capacity vs Temperature (d) Co-efficient of volume thermal expansion vs Temperature.

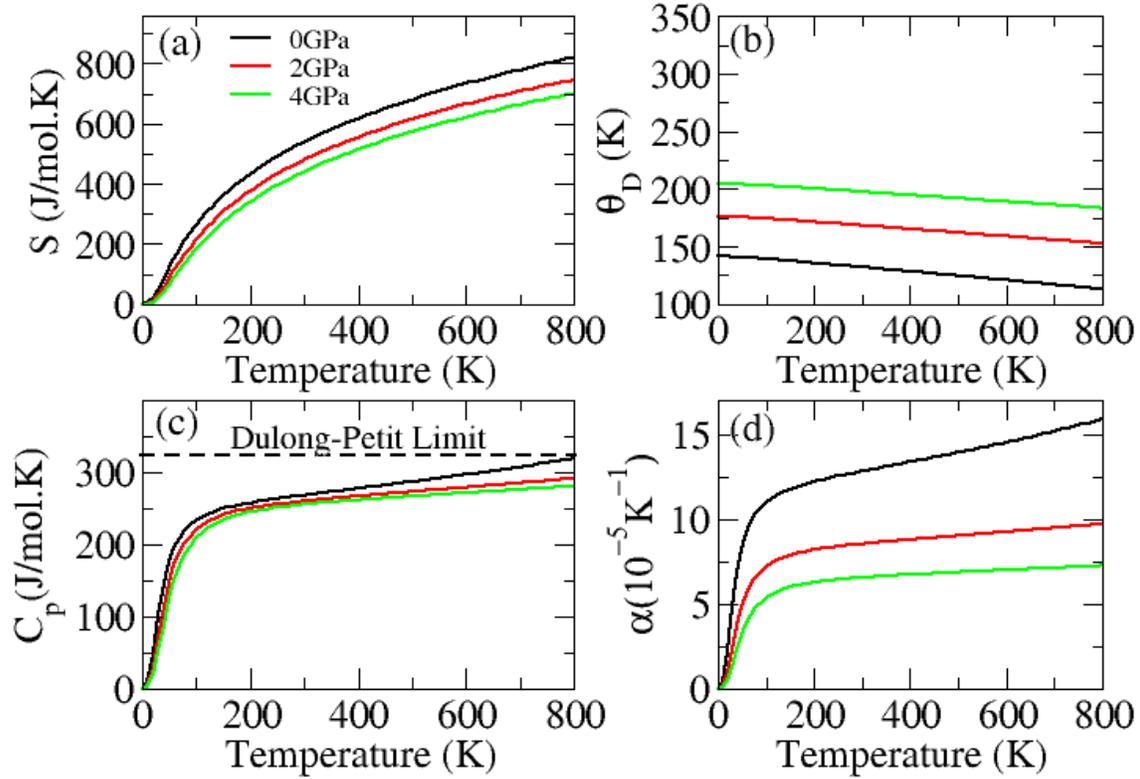

Figure 12: Calculated Thermodynamic properties of Li$_2$InBiI$_6$ (a) Entropy vs Temperature (b) Debye Temperature vs Temperature (c) Heat capacity vs Temperature (d) Co-efficient of volume thermal expansion vs Temperature.

## 4. Conclusion

This study employs first-principle calculations based on DFT to explore the crystallographic structure, formation energies, optical and electronic features of Li$_2$InBiX$_6$ (X= Cl, Br, I). The computational findings reveal there is a progressive decrease in the band gap as Cl is successively replaced by Br and I. Specific Heat and enthalpy were studied as thermo-dynamical properties. The extent of calculated bandgap lies from 1.1 eV to 1.7 eV of the studied perovskites, making them promising materials for a range of optoelectronic applications, especially solar cells. The analysis of the density of states and orbital-based band structure indicates that the valence band maxima is mainly originated by p-orbitals of X (Cl, Br and I), while the conduction band minima is governed by p-orbitals of Bi. Optical examination further reveals that the doping of halogens in perovskites enhances their optical absorption characteristics and thus improves its photoelectric conversion efficiency. The study emphasizes on the significant role of halogens substitution in the optical response and bandgap engineering. The assessment of thermoelectric outcome (especially, power factor) over the temperature range of 300 – 800K indicates improved thermal transport behaviour. The photo-catalytic properties makes it evident that $Li_2InBiCl_6$ showing for both oxidation and reduction of $H_2O$ while $Li_2InBiBr_6$ and $Li_2InBiI_6$ shows good response for both oxidation As a whole, the combination of favourable features-direct bandgap, strong absorption coefficient and high power factor highlights the potential of $Li_2InBiX_6$ (X= Cl, Br, I) for both thermoelectric and photovoltaic applications.

**Funding:**

The authors declare that no specific funding was received for this research